\documentclass[12pt]{iopart}

\usepackage{iopams}  
\usepackage{graphicx}
\usepackage{color}
\usepackage{mathrsfs}

\pdfoutput=1

\begin{document}

\title{Ultraslow scaled Brownian motion}

\author{Anna S. Bodrova$^{\dagger,\ddagger}$, Andrey G. Cherstvy$^\dagger$,
Aleksei V. Chechkin$^{\flat,\dagger}$, and Ralf Metzler$^{\dagger,\sharp}$}
\address{$\dagger$Institute of Physics and Astronomy, University of Potsdam, 14476
Potsdam, Germany\\
$\ddagger$Department of Physics, Moscow State University,  119899 Moscow, Russia\\
$\flat$Akhiezer Institute for Theoretical Physics, Kharkov Institute of Physics
and Technology, Kharkov 61108, Ukraine\\
$\sharp$Department of Physics, Tampere University of Technology, 33101 Tampere,
Finland}

\begin{abstract}
We define and study in detail \emph{utraslow scaled Brownian motion (USBM)\/}
characterised by a time dependent diffusion coefficient of the form $D(t)\simeq
1/t$. For unconfined motion the mean squared displacement (MSD) of USBM exhibits an
ultraslow, logarithmic growth as function of time, in contrast to the conventional
scaled Brownian motion. In an harmonic potential the MSD of USBM does not saturate
but asymptotically decays inverse-proportionally to time, reflecting the highly
non-stationary character of the process. We show that the process is weakly
non-ergodic in the sense that the time averaged MSD does not converge to the
regular MSD even at long times, and for unconfined motion combines a linear lag
time dependence with a logarithmic term. The weakly non-ergodic behaviour is
quantified in
terms of the ergodicity breaking parameter. The USBM process is also shown to be
ageing: observables of the system depend on the time gap between initiation of the
test particle and start of the measurement of its motion. Our analytical results
are shown to agree excellently with extensive computer simulations.
\end{abstract}

\section{Introduction}

In the wake of the development of modern particle tracking techniques strong
deviations of the time dependence of the mean squared displacement (MSD) from
the linear law $\langle x^2(t)\rangle\simeq t$ derived by Einstein \cite{einst}
and Smoluchowski \cite{smolu} have been observed in a variety of complex fluidic
environments \cite{hoefling,pt,igor_sm,pccp,meroz}. Typically, anomalous diffusion
of the power-law form
\begin{equation}
\label{msd_powerlaw}
\left< x^2(t)\right>\simeq t^\alpha
\end{equation}
is observed, where, depending on the value of the anomalous diffusion exponent
$\alpha$, we distinguish subdiffision with $0<\alpha<1$ and superdiffusion with
$\alpha>1$ \cite{bouchaud,report}. Accordingly, subdiffusion was observed in the
cytoplasm of living cells \cite{golding,tabei}, in artificially crowded liquids
\cite{lene1,weiss1}, and in structured or functionalised environments \cite{wong}.
Also superdiffusive motion was found in living cells \cite{christine,elbaum}.

Recently, interest in ultraslow diffusion processes with the logarithmic form
\begin{equation}
\label{msd}
\left< x^2(t)\right>\simeq\log^{\gamma}(t) 
\end{equation}
of the MSD with different values for the exponent $\gamma$ has been revived
\cite{pccp}. Ultraslow diffusion may be generated by periodically iterated
maps \cite{maps} and observed for random walks on bundled structures
\cite{bundled}. A prototype model for ultraslow diffusion is provided by
Sinai diffusion in quenched landscapes with random force field, for which
$\gamma=4$ \cite{sinai,dou1,gleb,aljaz}. In the context of Sinai diffusion
ultraslow continuous time random walks with super heavy-tailed
waiting times with $\gamma>0$ \cite{aljaz,havlin,chechkinepl,denisov1}
were discussed. Ultraslow scaling of the MSD of the form (\ref{msd}) were
obtained in aperiodic environments (variable $\gamma$)
\cite{aperiodic1} and vacancy induced motion ($\gamma=1$) \cite{oshanin}.
Moreover, it occurs in heterogeneous diffusion processes with exponentially
varying diffusivity ($\gamma=2$) \cite{HDP-PCCP}, or
interacting many-body systems in low dimensional disordered environments
with $\gamma=1/2$
\cite{sanders}, the dynamics of the latter being governed by an ultraslow,
ageing counting processes \cite{lomholt}.

The logarithmic time dependence (\ref{msd}) with $\gamma=1$ of the MSD is also
observed for the
self diffusion of particles in free cooling granular gases with constant, sub-unity
restitution coefficient in the homogeneous cooling state \cite{brilbook}. Granular
gases are rarefied granular systems, in which particles move along ballistic
trajectories between instantaneous collisions \cite{brilbook}. They are common in
Space, for instance, in protoplanetary discs, interstellar clouds and planetary
rings \cite{schmidtbook}. At terrestrial conditions granular gases may be obtained
by placing granular matter into containers with vibrating \cite{wildman} or rotating
\cite{rotdriv} walls. If no net external forces (gravitation, etc.) are acting on
the granular system, the motion of granular particles gradually slows down due to
dissipative collisions between them \cite{brilbook}. This microgravity condition
can be achieved, inter alia, with parabolic airplane flights or satellites
\cite{microgravity,flights,hayakawa} or by the use of diamagnetic levitation
\cite{magnetic}. We note that in very dense two-dimensional lattice gas systems,
ultraslow diffusion emerges, as well \cite{gleb_ultra}.

\begin{figure}
\begin{center}
\includegraphics[width=12cm]{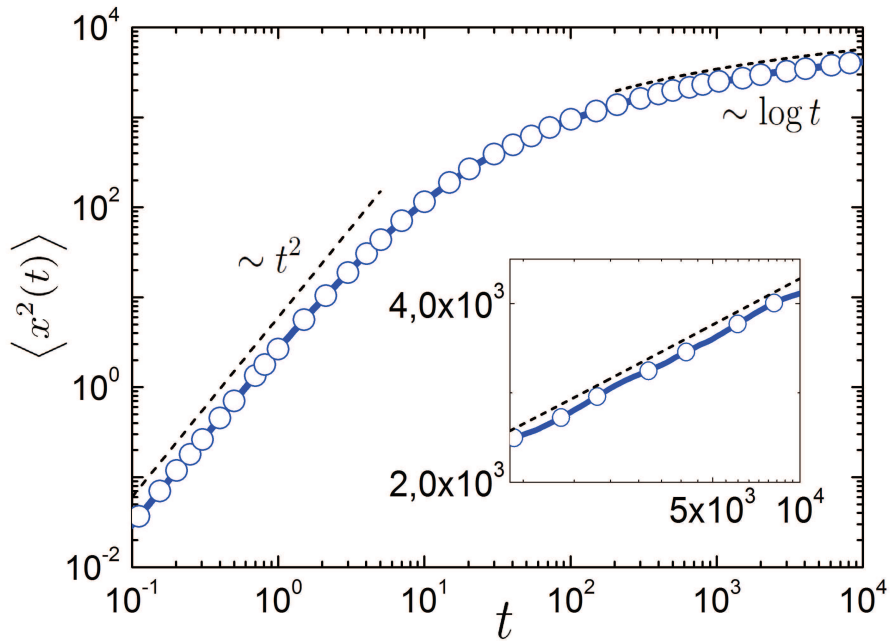}
\end{center}
\caption{Time dependence of the ensemble averaged MSD $\left< x^2(t)\right>$
obtained from event driven molecular dynamics simulations of three-dimensional
force-free granular gases \cite{annagg}. At short times the particles follow
ballistic trajectories, while for longer times the ensemble averaged MSD has
a logarithmic time dependence. The inset focuses on the logarithmic long time
behaviour.}
\label{Ggran}
\end{figure}

Figure \ref{Ggran} shows the crossover from the ballistic to the ultraslow form
(\ref{msd}) of the MSD of a granular gas with constant restitution coefficient
$\varepsilon=0.8$ in the homogeneous cooling state. Haff's law demonstrates that
the kinetic temperature of such a free granular gas with constant restitution
coefficient decays inverse-proportionally with time, $T(t)\simeq1/t$ \cite{haff}. 
For the effective self diffusion of the gas particles---mediated by
particle-particle collisions---this property translates into the time dependent
diffusion coefficient $D(t)\simeq1/t$ \cite{D1,D2,annagg}. We note that a diffusivity
of the form $D(t)=D_0+D_1/t$ with a component decaying inverse-proportionally with
time was used in the modelling of the motion of molecules in porous environments
\cite{Sen} as well as of water diffusion in brain tissue measured by magnetic
resonance imaging \cite{novikov}.

Here we study in detail the process of ultraslow scaled Brownian motion (USBM)
with time dependent diffusion coefficient $D(t)\simeq1/t$. Starting from the
Langevin equation for USBM and a summary of the simulations procedure we present
analytical and numerical results for the MSD and the time averaged MSD for the
cases of unconfined (Section \ref{free}) and confined (Section \ref{confined})
motion. We analyse in detail the disparity between the ensemble and time averaged
MSD and quantify the statistical scatter of the amplitude of the time averaged MSD
of individual realisations of the USBM process. Moreover we study the ageing
properties of USBM, that is, the explicit dependence of the physical observables
on the time difference between the initiation of the system and the start of the
observation. In Section \ref{concl} we present our Conclusions. In the Appendix
we present details of the calculation of higher order moments and the ergodicity
breaking parameter.

\section{Unconfined ultraslow scaled Brownian motion}
\label{free}

\subsection{Overdamped Langevin equation for ultraslow scaled Brownian motion}

\begin{figure}
\begin{center}
\includegraphics[width=12cm]{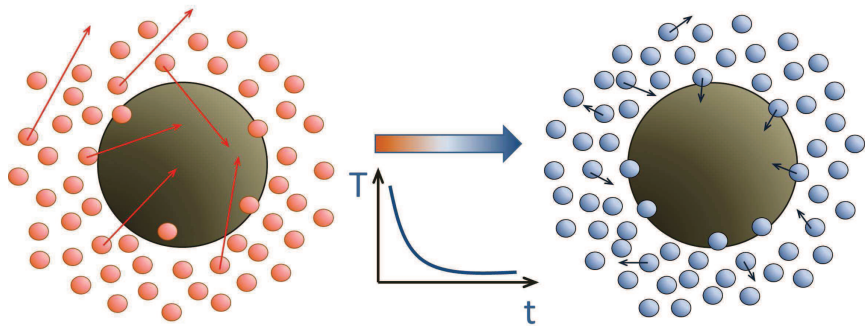}
\end{center}
\caption{Schematic of the motion of a Brownian particle in a bath with decreasing
temperature $T(t)\simeq t^{2\alpha-2}$ for $0\le\alpha<1$. The diffusion coefficient
of the Brownian particle decays with time as $D(t)\simeq t^{\alpha-1}$. USBM
corresponds to the case $\alpha=0$, while standard SBM is strictly limited to
$0<\alpha<2$ \cite{sbm}.}
\label{fig-scheme}
\end{figure}

Anomalous diffusion processes with power-law form (\ref{msd_powerlaw}) of the MSD
are often modelled in terms of scaled Brownian motion (SBM) characterised by an
explicitly time dependent diffusivity of the power-law form $D(t)\simeq t^{\alpha
-1}$ with $0<\alpha<2$, see, for instance, references \cite{weiss,verkman,wu,
szymaski,mitra,lutsko} as well as the study by Saxton \cite{saxton} and further
references therein. In SBM this form of $D(t)$ is combined with the regular
Langevin equation \cite{langevin}
\begin{equation}
\label{langevin}
\frac{dx(t)}{dt}=\sqrt{2D(t)}\times\zeta(t),
\end{equation}
in which $\zeta(t)$ represents white Gaussian noise with the normalised covariance
\begin{equation}
\left\langle \zeta(t_1)\zeta(t_2)\right\rangle=\delta(t_1-t_2)
\end{equation}
and zero mean $\langle\zeta(t)\rangle=0$. While for a system connected to a thermal
reservoir a description in terms of a time dependent temperature underlying SBM
is unphysical \cite{sbm}, time dependent diffusion coefficients appear naturally
in systems that are open or dissipate energy into other degrees of freedom such as
the granular gases discussed above, see the schematic in figure \ref{fig-scheme}.
In fact, granular gases with a viscoelastic, relative particle speed-dependent
restitution coefficient correspond to SBM with $\alpha=1/6$ \cite{brilbook,annagg}.
Diffusion in media with explicitly time dependent temperature can, for instance,
also be observed in snow melt dynamics \cite{molini,snow}.

A diffusion equation with a time dependent diffusivity proportional to $t^2$
was originally introduced by Batchelor \cite{batchelor} to describe the anomalous
Richardson relative diffusion \cite{richardson} in turbulent atmospheric systems.
SBM with diffusivity $D(t)\simeq t^{\alpha-1}$ was studied extensively during the
last few years \cite{LimSBM,fulinski,SokolovSBM,sbm,hadiseh}. In particular, the
weakly non-ergodic disparity between ensemble and time averages in SBM as well as
its ageing behaviour were analysed \cite{fulinski,SokolovSBM,sbm,hadiseh}, see
also below. Processes with both time and position dependent diffusion coefficients
were also reported \cite{andrey2015}. SBM is a Markovian process with stationary
increments $\zeta(t)$, however, it is rendered non-stationary by the time dependence
of the coefficient $D(t)$. SBM is therefore fundamentally different \cite{pccp,sbm}
from seemingly similar processes such as fractional Brownian motion or fractional
Langevin equation motion \cite{fbm}.

Following the motivation from our studies of granular gases with constant
restitution coefficient \cite{annagg} we here consider USBM with the time
dependent diffusion coefficient
\begin{equation}
\label{eq-dc-marginal-sbm}
D(t)=\frac{D_0}{1+t/\tau_0}.
\end{equation}
The time scale $\tau_0$ defines the characteristic time beyond which the long
time scaling $D(t)\sim D_0\tau_0/t$ sets in. We here introduce $\tau_0$ to avoid
a divergence of $D(t)$ at $t=0$. The case (\ref{eq-dc-marginal-sbm}) is explicitly
excluded in the allowed range for the scaling exponent $\alpha$ in SBM and, as we
will see, constitutes a new class of stochastic processes. In the following we solve
the overdamped Langevin equation (\ref{langevin}) with the time dependent diffusion
coefficient (\ref{eq-dc-marginal-sbm}) analytically and perform extensive computer
simulations of the corresponding finite-difference analogue of the Langevin equation.
In this procedure, at each time step the increment of the particle position takes
on the value
\begin{equation}
x_{i+1}-x_i=\sqrt{2D(i)}(W_{i+1}-W_i),\,\,\, i=0,1,2,\ldots,
\end{equation}
where $W_{i+1}-W_i$ is the increment of the standard Wiener process and $D(i)$ is
the value of the time dependent diffusivity (\ref{eq-dc-marginal-sbm}) at the time
instant $i$. We simulated $N=10^3$ independent particles (runs) with the parameters
$\tau_0=1$ and $D_0=1/2$ in all graphs presented below.

\subsection{Ensemble and time averaged mean squared displacements}
\label{sec-tamsd}

From direct integration of the Langevin equation (\ref{langevin}) with the time
dependent diffusivity  (\ref{eq-dc-marginal-sbm}) we find the ultraslow, logarithmic
growth
\begin{eqnarray}
\nonumber
\left\langle x^2(t) \right\rangle&=&\int_0^t\int_0^t\sqrt{D(t')D(t'')}\langle\zeta(t')
\zeta(t'')\rangle dt''dt'\\
&=&2D_0\tau_0\log\left(1+\frac{t}{\tau_0}\right)
\label{x2SBM}
\end{eqnarray}
of the ensemble averaged MSD. USBM therefore reproduces the asymptotic behaviour of
the MSD for granular gases in the homogeneous cooling state and with constant
restitution coefficient \cite{annagg}, as shown in figure \ref{Ggran}.

In addition to the ensemble averaged MSD $\langle x^2(t)\rangle$ of the particle
motion, it is often useful to compute the time averaged MSD
\begin{equation}
\overline{\delta^2(\Delta)}=\frac{1}{t-\Delta}\int_0^{t-\Delta}\Big[x(t'+\Delta)
-x(t')\Big]^2dt'.
\label{delta2} 
\end{equation}
Here, the lag time $\Delta$ defines the width of the averaging window slid over the
time series $x(t)$ of the particle position of overall length $t$ (the measurement
time). Time averages of the form (\ref{delta2}) are often used in experiments and
large scale simulations studies based on single particle tracking approaches, in
which typically few but long trajectories are available \cite{golding,tabei,weigel}.
The careful analysis of the time averaged MSD (\ref{delta2}) provides additional
important information on the studied process as compared to the ensemble averaged
MSD $\langle x^2(t)\rangle$, see, for instance, the analyses in references
\cite{tabei,weigel}. Often one takes the additional average over $N$ individual
particle traces $\overline{\delta^2_i(\Delta)}$,
\begin{equation}
\label{deltamean} 
\left<\overline{\delta^2(\Delta)}\right>=\frac{1}{N}\sum_{i=1}^{N}\overline{
\delta^2_i(\Delta)}.
\end{equation} 
For ergodic processes\footnote{We consider processes ergodic in the
Boltzmann-Khinchin sense when the long time average of a physical observable
converges to the associated time average.}
such as Brownian motion, fractional Brownian motion, and
fractional Langevin equation motion the time averaged MSD converges to the ensemble
averaged MSD in the limit of sufficiently long times, $\lim_{t\to\infty}\overline{
\delta^2(\Delta)}=\langle x^2(\Delta)\rangle$ \cite{pccp}. This property is due to
the stationarity of the increments of these processes \cite{yaglom}.
The ergodic behaviour $\lim_{t\to\infty}\overline{\delta^2(\Delta)}=\langle x^2(
\Delta)\rangle$ of these processes holds for unconfined motion when the system is
in fact out-of-equilibrium, an advantage of the particular definition
(\ref{delta2}). Moreover, ergodic systems fulfil the equivalence
\begin{equation}
\left<\overline{\delta^2(\Delta)}\right>=\left<x^2(\Delta)\right>
\end{equation}
even at finite $t$ \cite{pccp}. Systems in which we observe the disparity $\left<
\overline{\delta^2(\Delta)}\right>\neq\left<x^2(\Delta)\right>$ and therefore also
$\lim_{t\to\infty}\overline{\delta^2(\Delta)}\neq\left<x^2(\Delta)\right>$ are
called weakly non-ergodic \cite{pt,igor_sm,pccp,meroz,glass,pccp11}.\footnote{Note
that also transiently non-ergodic behaviour may become relevant as it may mask
intrinsic relaxation times when time averages are measured \cite{lene1,pre12}.}

\begin{figure}
\begin{center}
\includegraphics[width=12cm]{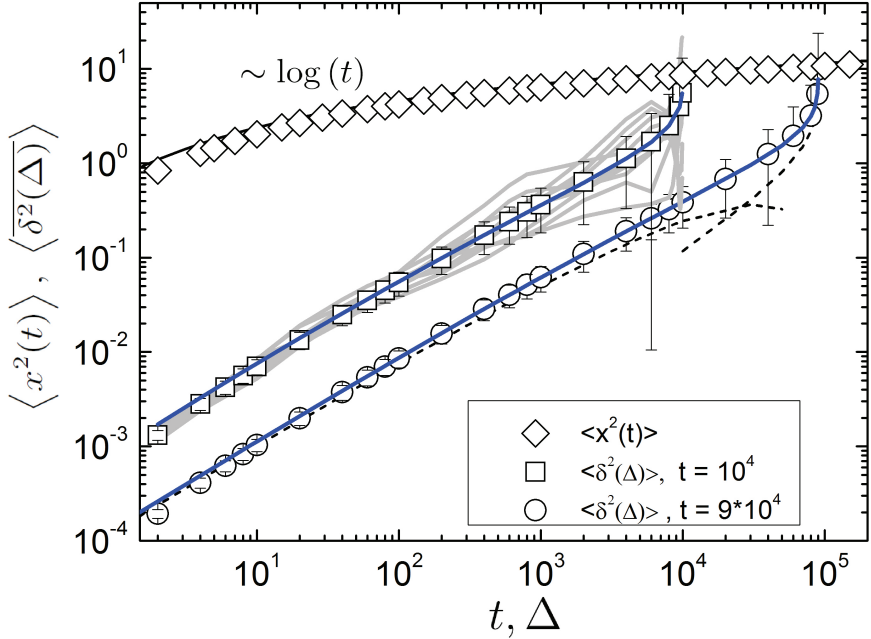}
\end{center}
\caption{Ensemble and time averaged MSDs for USBM with the time dependent diffusion
coefficient (\ref{eq-dc-marginal-sbm}). The analytical result (\ref{x2SBM}) for the
MSD $\left\langle x^2(t)\right\rangle$ shown by the black line compares nicely
with our simulations (diamonds). Similarly, the simulations results for different
measurement times (squares and circles) agree very well with the analytical result
(\ref{delta_full}) for the time averaged MSD $\left<\overline{\delta^2(\Delta)}
\right>$ shown by the blue lines for two different measurement times. The asymptotic
laws (\ref{d0}) and (\ref{d1}) are indicated by the dashed black lines. The thin
grey curves represent the results of the simulations for individual time traces.}
\label{Gdeltadamped}
\end{figure}

To calculate the time averaged MSD (\ref{deltamean}) for USBM we do not need to
consider the mixed position autocorrelations in the definition of the time averaged
MSD, as the expression in the angular brackets simplify as follows,
\begin{eqnarray}
\nonumber
\left<\overline{\delta^2(\Delta)}\right>&=&\frac{1}{t-\Delta}\int_0^{t-\Delta}
\left<\Big[x(t'+\Delta)-x(t')\Big]^2\right>dt'\\
&=&\frac{1}{t-\Delta}\int_0^{t-\Delta}\Big[\langle x^2(t'+\Delta)\rangle-\langle
x^2(t')\rangle\Big]dt'.
\end{eqnarray}
This is due to the property\footnote{In contrast, this is not valid in the case of
granular gases, where particles move ballistically in between instantaneous
collisions \cite{annagg}, or for processes driven by long-range correlated
increments such as fractional Brownian motion or fractional Langevin equation
motion \cite{pccp,fbm,BarkaiEB}.}
\begin{equation}
\label{xxx}
\langle x(t)x(t+\Delta)\rangle=\langle x^2(t)\rangle.
\end{equation}
for stochastic processes whose increments are independent random variables.
We thus find the exact form for the time averaged MSD of USBM,
\begin{equation}
\left<\overline{\delta^2(\Delta)}\right>=\frac{2D_0\tau_0}{t-\Delta}\Big[\ell(t)
-\ell(\Delta)-\ell(t-\Delta)\Big],
\label{delta_full}
\end{equation}
where we introduced the auxiliary function
\begin{equation}
\label{lt}
\ell(t)=(t+\tau_0)\log\left(1+\frac{t}{\tau_0}\right).
\end{equation}
The time averaged MSD (\ref{delta_full}) thus crosses over from the limiting
behaviour
\begin{equation}
\label{d0}
\left<\overline{\delta^2(\Delta)}\right>\sim2D_0\tau_0\frac{\Delta}{t}\log\left(
\frac{t}{\Delta}\right)
\end{equation}
at short lag times $\tau_0\ll\Delta\ll t$ combining a linear with a logarithmic
$\Delta$ dependence, to the purely logarithmic law
\begin{equation}
\label{d1}
\left<\overline{\delta^2(\Delta)}\right>\sim2D_0\tau_0\log\left(\frac{t+\tau_0}{t
-\Delta+\tau_0}\right)
\end{equation}
at $\tau_0\ll\Delta\approx t$. We see that as the lag time $\Delta$ approaches the
measurement time $t$, the time average MSD approaches the MSD (\ref{x2SBM}), $\left<
\overline{\delta^2(t)}\right>\to\left<x^2(t)\right>$. The results of our simulations
of the USBM process for both ensemble and time averaged MSDs agree very well
with the above analytical results, as demonstrated in figure \ref{Gdeltadamped}. In
that plot the thin grey curves depict the simulations results for the time averaged
MSD for individual trajectories. The amplitude spread between different trajectories
is fairly small for $\Delta\ll t$ and increases when the lag time $\Delta$
approaches the trace length $t$ due to worsening statistics. 

\subsection{Stochasticity of the time averaged mean squared displacement and
ergodicity breaking parameter}

\begin{figure}
\begin{center}
\includegraphics[width=12cm]{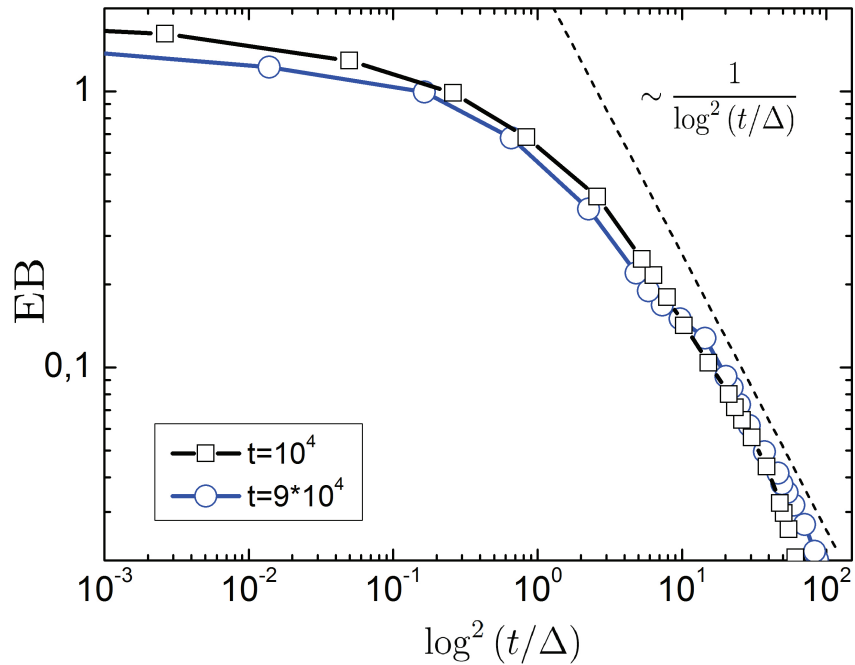}
\end{center} 
\caption{Ergodicity breaking parameter $\mathrm{EB}(\Delta)=\left\langle\xi^2(
\Delta)\right\rangle-1$ versus $\log^2\left(t/\Delta\right)$ for varying $\Delta$,
as obtained from computer simulations. The dashed line shows the asymptotic
(\ref{eq-eb-log}). Note the logarithm-squared horizontal axis.}
\label{GEB}
\end{figure}

Even ergodic processes such as Brownian motion exhibit a certain degree of
stochasticity of time averaged observables for shorter measurement times. The
amplitude fluctuations at a given lag time $\Delta$ of the time averaged MSD as
compared to the trajectory average (\ref{deltamean}) is quantified in terms of
the ergodicity breaking parameter \cite{pccp,BarkaiEB,he,rytovEB}
\begin{equation}
\label{EB0}
\mathrm{EB}(\Delta)=\lim_{t\to\infty}\frac{\left<\left(\overline{\delta^2(\Delta)}
\right)^2\right>-\left<\overline{\delta^2(\Delta)}\right>^2}{\left<\overline{\delta
^2(\Delta)}\right>^2}=\lim_{t\to\infty}\left<\xi^2\right>-1,
\end{equation}
where in the second equality we introduced the relative deviation \cite{he}
\begin{equation}
\xi=\frac{\overline{\delta^2(\Delta)}}{\left<\overline{\delta^2(\Delta)}\right>}.
\end{equation}
The necessary condition for ergodicity of a stochastic process is that the ergodicity
breaking parameter vanishes in the limit of infinitely long trajectories. Brownian
motion provides the basal level for the approach to ergodicity according to
\cite{BarkaiEB}
\begin{equation}
\mathrm{EB}_{\mathrm{BM}}=\frac{4\Delta}{3t}.
\label{EBM}
\end{equation}
Fractional Brownian motion and fractional Langevin equation motion are ergodic
\cite{fbm,BarkaiEB}. Weakly non-ergodic processes, which are characterised by the
disparity $\left<\overline{\delta^2(\Delta)}\right>\neq\langle x^2(\Delta)\rangle$
\cite{pt,igor_sm,pccp,meroz,pccp11,he} include continuous time random walks with
scale-free distributions of waiting times \cite{pt,igor_sm,pccp,pccp11,he} and
heterogeneous diffusion processes \cite{hdp,HDP-AGED}. In the limit of
long traces, the value of their ergodicity breaking parameter remains finite, which
is indicative of the intrinsic randomness of time averages of these processes. In
contrast, the ergodicity breaking parameter for SBM vanishes in the limit of
long trajectories \cite{SokolovSBM}. The ergodicity breaking parameter for USBM
is derived in the Appendix. The final expression in the relevant limit $\tau_0\ll
\Delta\ll t$ reads
\begin{equation}
\label{eq-eb-log}
\mathrm{EB}(\Delta)\sim\frac{4C}{\log^2\left(t/\Delta\right)}, 
\end{equation}
where the constant $C=\pi^2/6-1\simeq 0.645$. Thus, the time averaged MSD for USBM
becomes increasingly reproducible as the length of the time traces is extended,
albeit the approach to zero is logarithmically slow. We demonstrate the functional
form of the ergodicity breaking parameter as function of the lag time $\Delta$ for
two different measurement times and the approach of $\mathrm{EB}$ to its asymptotic
behaviour (\ref{eq-eb-log}) in figure \ref{GEB}.

\begin{figure}
\begin{center}
\includegraphics[width=12cm]{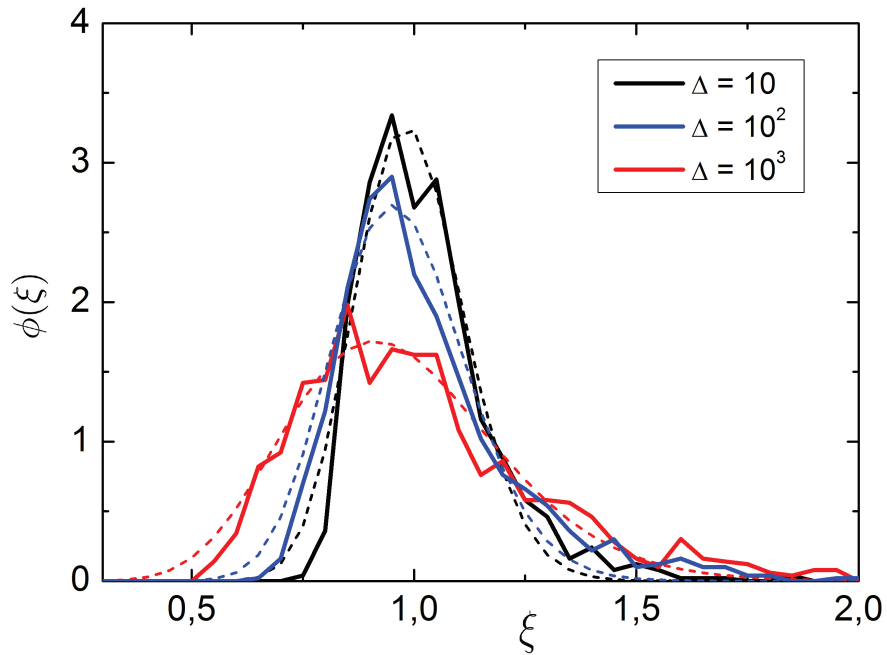}
\end{center}
\caption{Distribution $\phi(\xi)$ of the amplitude scatter of the time averaged
MSD. The dashed lines show the fit of the simulations data with the function $\phi
(\xi)\propto\exp(-a/\xi)\exp\left(-b\xi\right)$.}
\label{GPHI}
\end{figure}

The ergodicity breaking parameter quantifies the statistical spread of the time
averaged MSD. An important indicator for different types of stochastic processes
is also the complete distribution $\phi(\xi)$ \cite{pccp,pccp11,he,jpa}. As
shown in figure \ref{GPHI} this distribution has an asymmetric bell-shaped curve
approximately centred around the ergodic value $\xi=1$. The tail at larger
$\xi$ values appears somewhat longer compared to the tail at shorter
$\xi$.\footnote{For Brownian motion, fractional Brownian motion, and fractional
Langevin equation motion an approximately Gaussian shape of $\phi(\xi)$ is
found \cite{pccp,jpa}.} For longer lag times at fixed overall length $t$ of the
time series the width of the distribution $\phi(\xi)$ grows. This is consistent
with the fact that at larger value of $\Delta/t$ the time averages become more
random. In figure \ref{GPHI} we also show a fit to the function
\begin{equation}
\phi(\xi)\propto\exp(-a/\xi)\exp(-b\xi),
\end{equation}
which appears to capture the functional behaviour reasonably well. We note that
the shape of $\phi(\xi)$ appears narrower compared to the one of heterogeneous
diffusion processes with power-law space dependence of the diffusivity \cite{hdp}
which was fitted by a three-parameter Gamma distribution \cite{hdp,denis}.
In comparison, the distribution $\phi(\xi)$ for standard SBM is quite narrow,
although it widens as the exponent $\alpha$ approaches zero and particularly as
the lag time $\Delta$ grows \cite{sbm}.

\subsection{Ageing ultraslow scaled Brownian motion}

For processes with stationary increments such as Brownian motion or fractional
Brownian motion, if we initiate the system at $t=0$
but start recording it only at some later time $t_a$, the physical observables
will not explicitly depend on the ageing time $t_a$.\footnote{For confined
fractional Langevin equation motion, a transient ageing dependence exists 
\cite{jochen}.} However, for several anomalous processes pronounced ageing effects
are found. These include continuous time random walk processes with scale free
distributions of waiting times \cite{schulz,burov}, correlated continuous time
random walks \cite{tejedor}, non-linear maps generating
subdiffusion \cite{barkai2003}, systems with annealed and quenched disorder
\cite{kruesemann}, heterogeneous diffusion processes \cite{HDP-AGED}, or
standard SBM \cite{hadiseh}.

\begin{figure}
\begin{center}
\includegraphics[width=12cm]{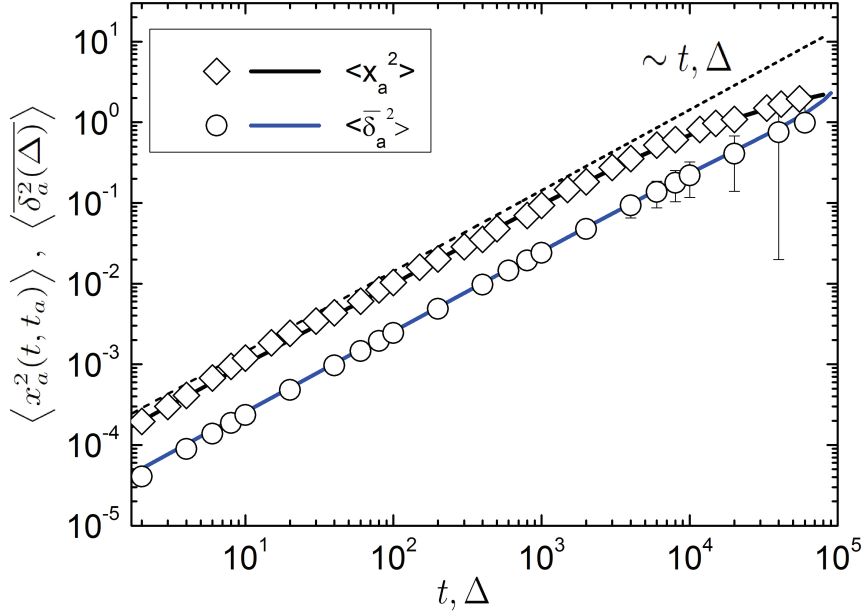}
\end{center}
\caption{Ensemble and time averaged MSDs $\langle x^2_a(t,t_a)\rangle$ and $\left<
\overline{\delta_a^2(\Delta)}\right>$ for ageing USBM. The measurement time is $t=9
\times10^4$ and the ageing time was chosen as $t_a=10^4$. Symbols: simulations
results. Lines: theoretical results of equations (\ref{x2SBMage}) and
(\ref{delta_age}).}
\label{Gage}
\end{figure}

In contrast to subdiffusive continuous time random walk processes, in which
ageing emerges due to the divergence of a characteristic waiting time \cite{schulz},
in ultraslow SBM the non-stationarity of the system stems from the explicit time
dependence of the diffusion coefficient. When the recording of the particle position
starts at a finite time $t_a$, this ageing time explicitly appears in the particle's
MSD. For the aged MSD \cite{schulz,pccp} in analogy to equation (\ref{x2SBM}) we find
that
\begin{eqnarray}
\nonumber
\left< x^2_a(t,t_a)\right>&=&2\int_{t_a}^{t_a+t}\int_{t_a}^{t_a+t}\left<\sqrt{D(t')}
\zeta(t')\sqrt{D(t'')}\zeta(t'')\right>dt'dt''\\
&=&2D_0\tau_0\log\left(1+\frac{t}{t_a+\tau_0}\right).
\label{x2SBMage}
\end{eqnarray}
In the limit of strong ageing, $ t_a\gg t$, this expression yields the linear
scaling
\begin{equation}
\left\langle x^2_a(t,t_a) \right\rangle \approx 2D_0\tau_0\frac{t}{t_a}.
\label{xata}
\end{equation}
of the MSD with time $t$, the ageing time $t_a$ rescaling the effective particle
diffusivity. The transition between this ageing-dominated linear scaling for the
MSD and the anomalous logarithmic time dependence in the weak ageing limit $t\gg
t_a$ is clearly seen in figure \ref{Gage}.

For the aged time averaged MSD \cite{pccp,schulz} we obtain the result
\begin{eqnarray}
\nonumber
\left<\overline{\delta_a^2(\Delta,t_{a})}\right>&=&\frac{1}{t-\Delta}\int_{t_a}^{
t-\Delta+t_a}\left<\Big[x(t'+\Delta)-x(t')\Big]^2\right>dt'\\
&=&\frac{2D_0\tau_0}{t-\Delta}\Big[\ell(t_a+t)-\ell(t_a+\Delta)-\ell(t_a+t-\Delta)
+\ell(t_a)\Big],
\label{delta_age}
\end{eqnarray}
where the auxiliary function $\ell(t)$ was defined in equation (\ref{lt}). In the
limit
$\tau_0\ll\Delta\ll t$ and $\Delta\ll t_a$ the aged time averaged MSD factories
into a term containing all the information on the ageing and measurement times $t_a$
and $t$, and another capturing the physically relevant dependence on the lag time
$\Delta$ and the measurement time $t$,
\begin{equation}
\left<\overline{\delta_a^2(\Delta,t_a)}\right>\sim2D_0\tau_0\frac{\Delta}{t}\log
\left(1+\frac{t}{t_a}\right).
\label{da}
\end{equation}
This factorisation is analogous to that of heterogeneous diffusion processes
\cite{HDP-AGED}, scale-free subdiffusive continuous time random walks \cite{schulz},
and standard SBM \cite{hadiseh}. However, in contrast to these processes the aged
time averaged MSD for short lag times does not factorise into the product of the
non-aged time averaged MSD (\ref{d1}) and a factor containing the ageing time.

\begin{figure}
\begin{center}
\includegraphics[width=12cm]{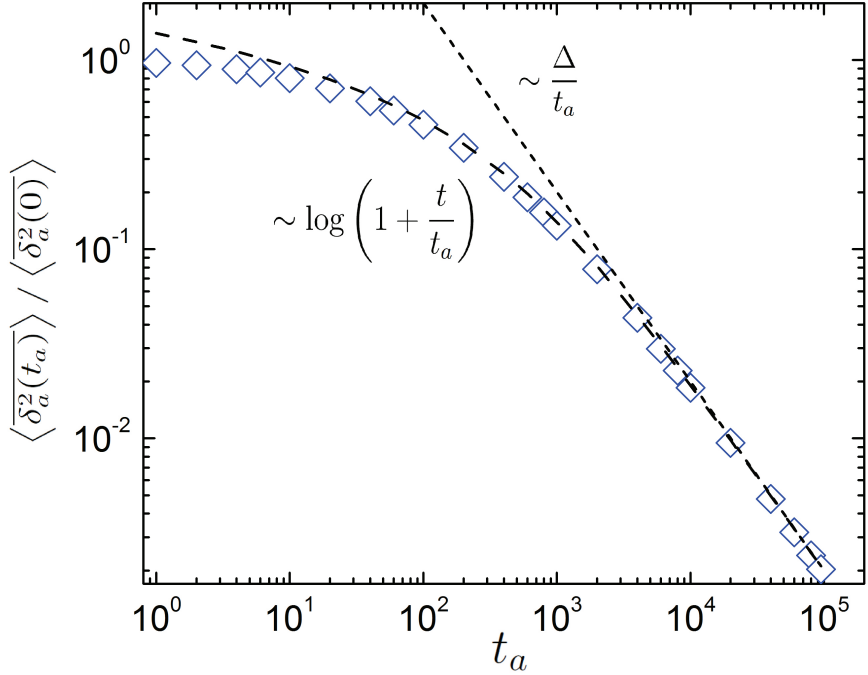}
\end{center}
\caption{Time averaged MSD $\left<\overline{\delta_a^2}\right>$ versus ageing time
$t_a$. The analytical results (\ref{da}) and (\ref{das}) are shown by the dashed
lines, while the symbols correspond to the results of simulations. Parameters:
measurement time $t=10^3$ and lag time $\Delta=10$.}
\label{Gandrey}
\end{figure}

For strong ageing $t_a\gg t$ we obtain the linear scaling
\begin{equation}
\left<\overline{\delta_a^2(\Delta,t_a)}\right>\sim2D_0\tau_0\frac{\Delta}{t_a}.
\label{das}
\end{equation}
In this limit, that is, the system becomes apparently ergodic and we observe the
equality $\left<\overline{\delta_a^2(\Delta,t_a)}\right>=\langle x^2_a(\Delta,t_a)
\rangle$, as can be seen from comparison with equations (\ref{xata}) and
(\ref{das}). Figure \ref{Gandrey} shows the convergence of the time averaged MSD
to the limiting behaviour (\ref{das}).
Such a behaviour was previously observed for aged subdiffusive SBM \cite{hadiseh},
heterogeneous diffusion processes \cite{HDP-AGED}, and continuous time random walk
processes \cite{schulz}. In the case of USBM this phenomena has a clear physical
explanation: at the beginning of the experiment the diffusion coefficient $D(t)$
significantly decreases during the measurement time $t\gg\tau_0$ from $D(0)=D_0$
to $D(t)\sim D_0\tau_0/t$, and the system is strongly non-stationary. In contrast,
after a long ageing period $t_a\gg t$ the diffusion coefficient remains practically
unchanged during the measurement time, $D(t_a+t)\simeq D(t_a)=D_0\tau_0/t_a$. 

\begin{figure}
\begin{center}
\includegraphics[width=12cm]{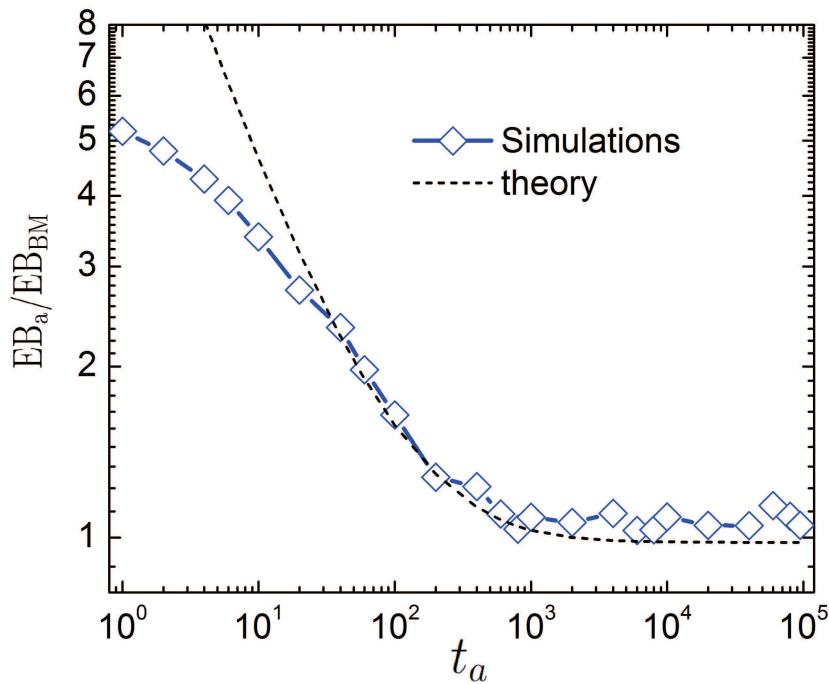}
\end{center}
\caption{Ergodicity breaking parameter $\mathrm{EB}_a$ normalised by the Brownian
value ${\mathrm{EB}_{\rm BM}}$ (\ref{EBM}) as function of ageing time $t_a$. The
dashed lines correspond to the analytical result (\ref{EBa}), while the symbols are
the results of simulations. Parameters are the same as in figure~\ref{Gandrey}.
For strong ageing, $t_a\gg t$, $\mathrm{EB}_a$ does not depend on $t_a$ and
approaches ${\mathrm{EB}_{\rm BM}}$ (\ref{EBM}).}
\label{GEBa}
\end{figure}

Figure \ref{Gandrey} explicitly shows how
the amplitude of the time averaged MSD is reduced due to ageing in the system.
How do the fluctuations of individual time averaged MSD traces change in the
presence of ageing? The derivation of the ergodicity breaking parameter for the
aged process is provided in the Appendix. The final result in the limit $\Delta\ll
t$ and $\Delta\ll t_a$ assumes the form
\begin{equation}
\mathrm{EB}_a=\frac{4\Delta t/t_a}{3t_a(1+t/t_a)\log^2(1+t/t_a)}.
\label{EBa}
\end{equation}
In the strong ageing limit $t_a\gg t$ the ergodicity breaking parameter $\mathrm{EB}
_a$ is independent of the ageing time $t_a$, and it asymptotically converges to
the result (\ref{EBM}) of Brownian diffusion. Our theoretical results agree well
with the simulations, as witnessed by figure \ref{GEBa}. For weak ageing $t_a\ll
\Delta,t$ the result (\ref{delta_full}) of the non-aged USBM process is recovered.

\section{Confined ultraslow scaled Brownian motion}
\label{confined}

The motion of particles in external confinement is an important physical concept
for applications of stochastic processes, and it is also relevant from an 
experimental point of view. Namely, the motion of particles in cells may repeatedly
hit the cell wall, or the tracer particles may experience a restoring force in
particle tracing experiments by help of optical tweezers. Here we consider the
generic case of confinement in an harmonic potential. USBM in the presence of such
a linear restoring force is governed by the overdamped Langevin equation with
additional Hookean force term $-kx$,
\begin{equation}
\frac{dx}{dt}=\sqrt{2D(t)}\times\zeta(t)-kx.
\label{DampedLangevinConf}
\end{equation}

\subsection{Ensemble and time averaged mean squared displacements}

The ensemble averaged MSD follows directly from this stochastic equation, and we
obtain
\begin{equation}
\left\langle x^2(t) \right\rangle = 2D_0\tau_0\mathscr{E}(t+\tau_0).
\label{x2conf}
\end{equation}
Here we defined the auxiliary function
\begin{eqnarray}
\nonumber
\mathscr{E}(x)&=&e^{-2kx}\int_{2k\tau_0}^{2kx}\frac{\exp(-y)}{y}dy\\
&=&e^{-2k\left(x+\tau_0\right)}\Big[\mathrm{Ei}(2kx)-\mathrm{Ei}(2k\tau_0)\Big]
\label{mathcal}
\end{eqnarray}
where in the second line we used the definition of the exponential integral
\begin{equation}
\mathrm{Ei}(z)=-\int_{-z}^{\infty}\frac{\exp(-y)}{y}dy.
\end{equation}
The asymptotic behaviour of the MSD for long times $t\gg 1/k$ has the time
dependence
\begin{equation}
\langle x^2(t)\rangle=\frac{D_0\tau_0}{kt}.
\label{x2confas}
\end{equation}
Reflecting the temporal decay of the temperature encoded in the time dependent
diffusion coefficient (\ref{eq-dc-marginal-sbm}) we observe the $1/t$ scaling of
the MSD in confinement. This underlines the highly non-stationary and athermal
character of this process \cite{fulinski,sbm,hadiseh}.

\begin{figure}
\begin{center}
\includegraphics[width=12cm]{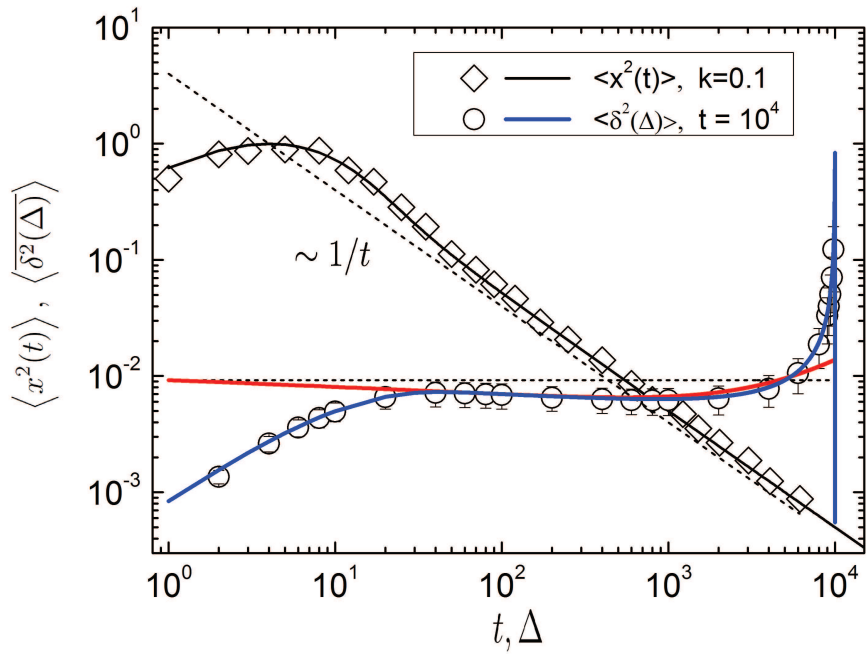}
\end{center}
\caption{Ensemble and time averaged MSDs $\langle x^2(t)\rangle$ and $\left<
\overline{\delta^2(\Delta)}\right>$ for confined USBM. The black line represents
the analytical result (\ref{x2conf}), while the blue line denotes equation
(\ref{dconf}). The red line shows the asymptotic behaviour (\ref{dconfas}),
and the horizontal dashed line the leading term (\ref{dconfmean}).
The symbols correspond to the simulations of equation
(\ref{DampedLangevinConf}).}
\label{Gconf}
\end{figure}

The time averaged MSD for confined USBM is obtained from the relation
\begin{equation}
\label{dddconf}
\left<\overline{\delta^2(\Delta)}\right>=\frac{1}{t-\Delta}\int_0^{t-\Delta}\Big[
\langle x^2(t'+\Delta)\rangle-2\langle x(t')x(t'+\Delta)\rangle+\langle x^2(t')
\rangle\Big].
\end{equation}
The covariance of the position for ultraslow SBM in confinement can no longer be
simplified according to equation (\ref{xxx}) but has the time dependence
\begin{equation}
\label{xcor}
\langle x(t_1)x(t_2)\rangle=2D_0\tau_0e^{-k\left(t_2-t_1\right)}\mathscr{E}(\tau_0
+t_1).
\end{equation}
Introducing relations (\ref{x2conf}) and (\ref{xcor}) into equation (\ref{dddconf})
we obtain
\begin{eqnarray}
\nonumber
\left<\overline{\delta^2(\Delta)}\right>&=&\frac{D_0\tau_0}{(t-\Delta)k}\Bigg\{
\log\left(\frac{t+\tau_0}{\Delta+\tau_0}\right)-\mathscr{E}(t+\tau_0)+\\
\nonumber
&&+\left(1-2e^{-k\Delta}\right)\left[\log\left(1+\frac{t-\Delta}{\tau_0}\right)
-\mathscr{E}(t-\Delta+\tau_0)\right]\\
&&+\mathscr{E}(\Delta+\tau_0)\Bigg\}.
\label{dconf}
\end{eqnarray}
For long times and strong external confinement, $\left\{t,t_a,\Delta\right\}\gg
\left\{1/k,\tau_0\right\}$ this expression simplifies to
\begin{equation}
\left<\overline{\delta^2(\Delta)}\right>=\frac{D_0\tau_0}{kt}\left[2\left(1+\frac{
\Delta}{t}\right)\log\left(\frac{t}{\tau_0}\right)-\log\left(\frac{\Delta}{\tau_0}
\right)\right].
\label{dconfas}
\end{equation}
The time averaged MSD has a pronounced plateau for $\Delta\ll t,t_a$,
\begin{equation}
\left<\overline{\delta^2(\Delta)}\right>=\frac{2D_0\tau_0}{kt}\log\left(\frac{t}{
\tau_0}\right),
\label{dconfmean}
\end{equation}
that is, in this regime the time averaged MSD is independent of the lag time,
compare the discussion in references \cite{sbm,hadiseh}.
Simulations based on the Langevin equation with the Hookean forcing are in excellent
agreement with these analytical results, as shown in figure \ref{Gconf}.

\subsection{Ageing ultraslow scaled Brownian motion in confinement}

\subsubsection{Ensemble averaged mean squared displacement.}

\begin{figure}
\begin{center}
\includegraphics[width=12cm]{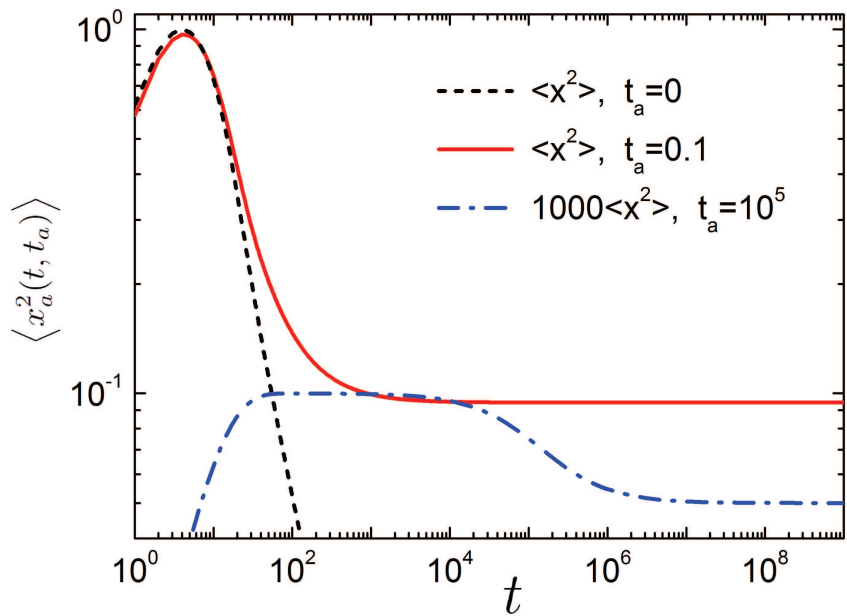}
\end{center}
\caption{Ensemble averaged MSD $\langle x_a^2(t,t_a)\rangle$ for confined ageing
USBM at different ageing times: $t_a=0$ (no ageing, black line), $t_a=0.1$ (weak
ageing, red line), and $t_a=10^5$ (strong ageing, blue line). Note that for
better visibility the curve for $t_a=10^5$ was multiplied by a factor of $10^3$.}
\label{Gxconfage}
\end{figure}

For confined ageing USBM, in which we measure the MSD starting from the ageing
time $t_a$ until time $t$, the result for the MSD becomes
\begin{eqnarray}
\nonumber
\left< x_a^2(t,t_a)\right>&=&\left<\left[x(t_a+t)-x(t_a)\right]^2\right>\\
\nonumber
&=&\left< x^2(t_a+t)\right>+\left< x^2(t_a)\right>-2\left< x(t_a+t)x(t_a)\right>\\
&=&2D_0\tau_0\Big[\mathscr{E}(t_a+\tau_0)+\mathscr{E}(t_a+t+\tau_0)-2e^{-kt}
\mathscr{E}(t_a+\tau_0)\Big],
\label{x2SBMa}
\end{eqnarray}
where $\mathscr{E}(x)$ is defined in equation (\ref{mathcal}). Expression
(\ref{x2SBMa}) reduces to equation (\ref{x2conf}) for vanishing ageing, $t_a=0$.
However, even in the presence of weak ageing, $t_a\ll1/k$, at long times $t\gg1/k$
the behaviour of the MSD reads
\begin{equation}
\left< x_a^2(t,t_a)\right>=2D_0\tau_0\log\left(1+\frac{t_a}{\tau_0}\right)+\frac{
D_0\tau_0}{kt}.
\label{xconfage}
\end{equation}
contrasting the behaviour in equation (\ref{x2confas}). The ensemble averaged MSD
for ageing USBM at different ageing times is depicted in
figure \ref{Gxconfage}. At short times $t<1/k$ the weakly aged MSD follows the
non-aged behaviour. Eventually it attains the plateau given by the first term in
equation (\ref{xconfage}), instead of decaying towards zero as in the non-aged case.
In the analysis of experimental data times the exact moment of the system's
initiation may often not be known, for instance, when
measuring biological cells. The apparent plateau revealed here for confined ageing
USBM dynamics may thus erroneously be mistaken as a signature of a stationary
process.

Expanding the exponential integral in equation (\ref{x2SBMa}), in the strong ageing
limit $t_a\gg \left\{\tau_0,1/k\right\}$ we find
\begin{equation}
\langle x_a^2(t,t_a)\rangle=\frac{D_0\tau_0}{kt_a}\left(1+\frac{1}{1+t/t_a}-2e^{
-kt}\right).
\label{xage}
\end{equation}
For $t\ll 1/k$ we recover the unconfined result (\ref{xata}). In the opposite
limit $t\gg 1/k$ the behaviour of equation (\ref{xage}) crosses over to
\begin{equation}
\label{xctat}
\langle x_a^2(t,t_a)\rangle=\frac{D_0\tau_0}{k}\left(\frac{1}{t_a}+\frac{1}{t_a+t}
\right).
\end{equation}
In this case we recover a transition between two plateaus, as it was observed for
subdiffusive SBM \cite{hadiseh}. Namely, for short measurement times $t\ll t_a$ we
find from result (\ref{xctat}) that
\begin{equation}
\label{short}
\langle x_a^2(t,t_a)\rangle=\frac{2D_0\tau_0}{kt_a},
\end{equation}
while at long measurement times $t\gg t_a$ this turns to
\begin{equation}
\label{long}
\langle x_a^2(t,t_a)\rangle=\frac{D_0\tau_0}{kt_a}.
\end{equation}
This behaviour, which appears unique for USBM and SBM, is depicted in
figure \ref{Gxconfage}).

\begin{figure}
\begin{center}
\includegraphics[width=12cm]{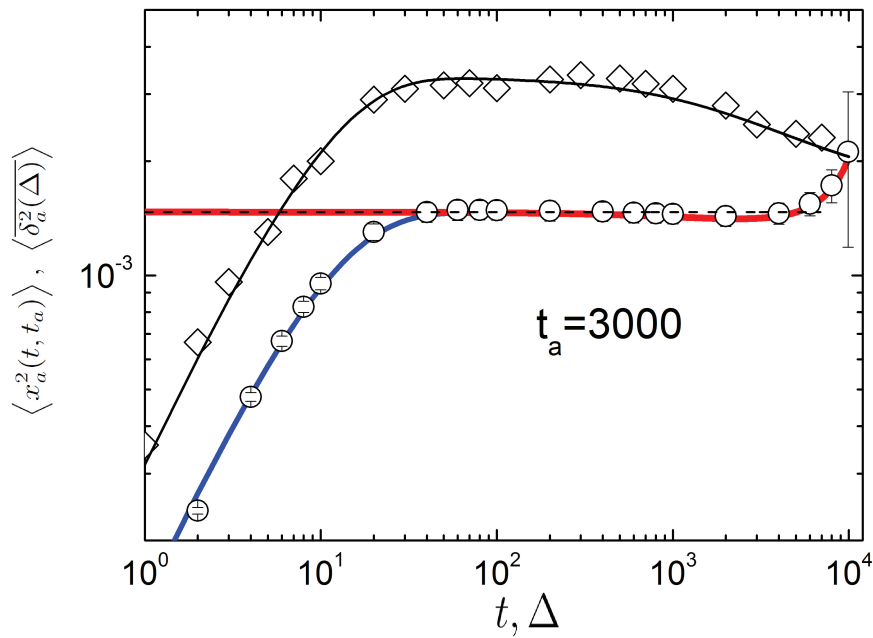}
\end{center}
\caption{Ensemble and time averaged MSDs $\langle x_a^2(t,t_a)\rangle$ and $\left<
\overline{\delta_a^2(\Delta)}\right>$ for confined ageing USBM. The symbols depict
simulations of equation (\ref{DampedLangevinConf}). The blue line corresponds to the
theoretical result (\ref{dconfage}), and the red line shows the asymptotic
(\ref{dconfagas}). The horizontal dashed line shows the leading term
(\ref{dconfagasas}).}
\label{Gconfage}
\end{figure}

\subsubsection{Time averaged mean squared displacement.}

The time averaged MSD for ageing confined USBM is derived analogously to the
non-aged case, yielding
\begin{eqnarray}
\nonumber
\left<\overline{\delta_a^2(\Delta,t_a)}\right>&=&\frac{D_0\tau_0}{(t-\Delta)k}
\left\{\left(1-2e^{-k\Delta}\right)\left[\log\left(1+\frac{t-\Delta}{t_a+\tau_0}
\right)\right.\right.\\
\nonumber
&&\hspace*{1.8cm}-\mathscr{E}(t_a+t-\Delta+\tau_0)+\mathscr{E}(t_a+\tau_0)\Bigg]\\
\nonumber
&&+\log\left(\frac{t_a+t+\tau_0}{t_a+\Delta+\tau_0}\right)\\
&&-\mathscr{E}(t_a+t+\tau_0)+\mathscr{E}(t_a+\Delta+\tau_0)\Bigg\}.
\label{dconfage}
\end{eqnarray}
In the limit of strong confinement $1/k\ll\left\{t_a,t,\Delta\right\}$ this
expression can be significantly simplified to obtain
\begin{equation}
\left<\overline{\delta_a^2(\Delta,t_a)}\right>=\frac{D_0\tau_0}{k(t-\Delta)}\left[
\log\left(\frac{t+t_a+\tau_0}{t_a+\Delta+\tau_0}\right)+\log\left(1+\frac{t-\Delta
}{t_a+\tau_0}\right)\right].
\label{dconfagas}
\end{equation}
For $\Delta\ll t,t_a$ we again find an apparent plateau,
\begin{equation}
\left<\overline{\delta_a^2(\Delta,t_a)}\right>=\frac{2D_0\tau_0}{kt}\log\left(1+
\frac{t}{t+t_a}\right).
\label{dconfagasas}
\end{equation}
In the case of strong ageing $t_a\gg t$ we find
\begin{equation}
\left<\overline{\delta_a^2(\Delta,t_a)}\right>=\frac{2D_0\tau_0}{kt_a}.
\end{equation}
Comparison to equation (\ref{short}) shows that the time averaged MSD becomes equal
to the ensemble MSD in this strong ageing regime, and ergodicity is apparently
restored as in the unconfined case. The behaviour of the ensemble and time
averaged MSDs for confined ageing USBM are depicted in figure \ref{Gconfage}.

\begin{figure}
\begin{center}
\includegraphics[width=12cm]{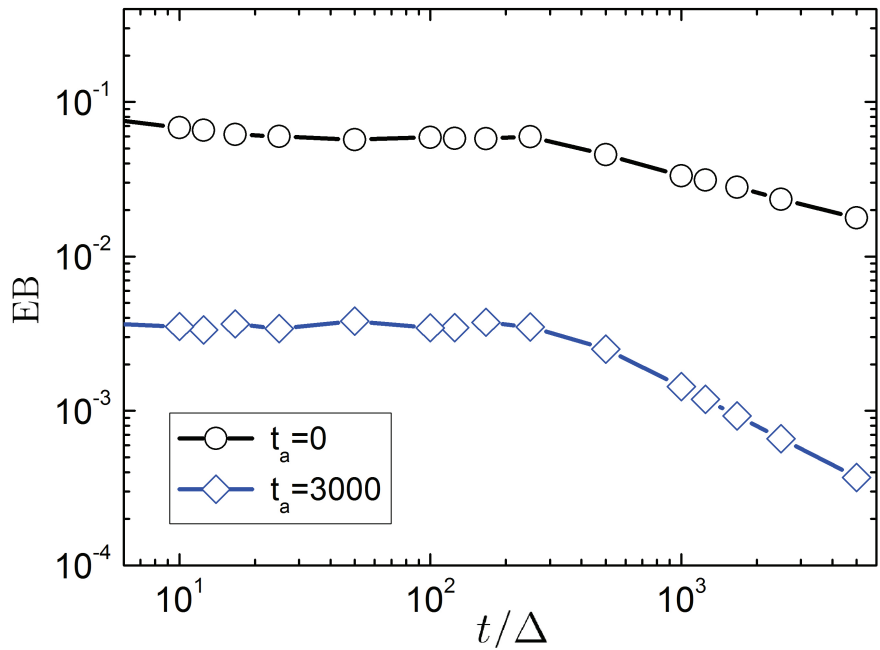}
\end{center}
\caption{Ergodicity breaking parameter $\mathrm{EB}$ as function of $t/\Delta$ in
the non-aged ($t_a=0$) and aged ($t_a=3000$) cases.}
\label{GEBconf}
\end{figure}

The ergodicity breaking parameter $\mathrm{EB}$ for confined USBM is depicted in
figure \ref{GEBconf} for both absence and presence of ageing. It is a decreasing
function of the ratio $t/\Delta$ for large $t/\Delta$, while at small values of
$t/\Delta$ it remains practically unchanged.

\section{Conclusions}
\label{concl}

We proposed and studied ultraslow scaled Brownian motion, a new anomalous
stochastic process with a time dependent diffusion coefficient of the form
$D(t)\simeq1/t$. Formally USBM corresponds to the lower bound $\alpha=0$
of scaled Brownian motion with diffusivity $D(t)\simeq t^{\alpha-1}$ ($0<\alpha$)
\cite{LimSBM,fulinski,sbm,SokolovSBM,hadiseh}, yet its dynamical behaviour is
significantly different. We showed that USBM yields a
logarithmic time dependence of the MSD rather than the power-law scaling of SBM.
USBM's time averaged MSD was shown to acquire a combination of power-law and
logarithmic lag time dependence. USBM is weakly non-ergodic and ageing. The
ergodicity breaking
parameter quantifying the random character of time averages of the MSD has a weak
logarithmic dependence on the ratio $\Delta/t$ of lag time $\Delta$ and length
$t$ of the recorded trajectories, tending to zero in the limit of infinitely long
traces and/or short lag times. In the case of strong ageing the system tends to
usual Brownian motion and the behaviour of the system becomes apparently ergodic.
Under external confinement the behaviour of the USBM dynamics exhibits an apparent
plateau for the time averaged MSD, while the ensemble averaged MSD decays
proportionally to $1/t$ at longer times, reflecting the highly non-stationary
character of USBM. Ageing produces an apparent plateau for the ensemble averaged
MSD and a crossover between two plateaus for the time averaged MSD. USBM adds to
the rich variety of ultraslow processes with logarithmic growth of the ensemble
averaged MSD yet displays several unique features in comparison to other
ultraslow processes.

Potential applications of USBM are foremost in the description of random particle
motion in intrinsically non-equilibrium system such as free cooling granular gases
or systems coupled to explicitly time dependent thermal reservoirs. On a more
general level we hope that the discussion of ultraslow processes will lead to a
rethinking of claims in diffusion studies that certain particles appear immobile.
Namely, one often observes a population splitting into a (growing) fraction of
immobile particles and another fraction of particles performing anomalous diffusion
of the form (\ref{msd_powerlaw}) \cite{pop_split}. Ageing continuous time random
walks \cite{schulz} or heterogeneous diffusion processes \cite{HDP-PCCP,HDP-AGED}
give rise to such a behaviour. However, given the tools provided here on ultraslow
diffusion it might be worthwhile checking whether the observe ``immobile''
particles may in fact perform logarithmically slow diffusion.

\ack

The authors thank N. V. Brilliantov, A. Godec, I. M. Sokolov and F. Spahn
for stimulating discussions. The simulations were run at the Chebyshev
supercomputer of the Moscow State University. This work was supported by
the EU IRSES DCP-PhysBio N269139 project, the Academy of Finland (FiDiPro
scheme to RM), Berlin Mathematical Society (to AVC) and the Deutsche
Forschungsgemeinschaft (DFG Grant CH 707/5-1 to AGC).

\appendix

\section{Derivation of the ergodicity breaking parameter}

The ergodicity breaking parameter (\ref{EB0}) requires the fourth order moment
\begin{eqnarray}
\nonumber
\left<\left(\overline{\delta^2(\Delta)}\right)^2\right>&=&\frac{1}{\left(t-\Delta
\right)^2}\int_0^{t-\Delta}\int_0^{t-\Delta}\left<\left(x(t_1+\Delta)-x(t_1)
\right)^2\right.\\
&&\hspace*{3.2cm}\times\left.\left(x(t_2+\Delta)-x(t_2)\right)^2\right> dt_2dt_1.
\end{eqnarray}
Using Isserlis' or Wick's theorem the integrand can be rewritten in the form
\begin{eqnarray}
\nonumber
\langle\left(x(t_1+\Delta)-x(t_1)\right)^2\left(x(t_2+\Delta)-x(t_2)\right)^2
\rangle\\
\nonumber
&&\hspace*{-7.2cm}
=\left\langle\left(x(t_1+\Delta)-x(t_1)\right)^2\right\rangle\left\langle\left(x(
t_2+\Delta)-x(t_2)\right)^2\right\rangle\\
&&\hspace*{-6.8cm}
+2\left\langle\left(x(t_1+\Delta)-x(t_1)\right)\left(x(t_2+\Delta)-x(t_2)\right)
\right\rangle^2.
\end{eqnarray}
The numerator in equation (\ref{EB0}) may thus be represented as
\begin{eqnarray}
\nonumber
\mathcal{N}&=&\left<\left(\overline{\delta^2(\Delta)}\right)^2\right>-\left<
\overline{\delta^2(\Delta)}\right>^2\\
\nonumber
&=&\frac{2}{\left(t-\Delta\right)^2}\int_0^{t-\Delta}\int_0^{t-\Delta}\left<[
x(t_1+\Delta)-x(t_1)]\right.\\
&&\hspace*{4cm}\left.\times[x(t_2+\Delta)-x(t_2)]\right>^2dt_2dt_1.
\label{a1}
\end{eqnarray}
Taking into account relation (\ref{xxx}) and the symmetry of expression (\ref{a1})
with respect to $t_1$ and $t_2$, we get
\begin{eqnarray}
\nonumber
\mathcal{N}&=&\frac{4}{\left(t-\Delta\right)^2}\\
&&\times\int_0^{t-\Delta}\int_{t_1}^{t-
\Delta}\left(\left\langle x^2\left(t_1+\Delta\right)\right\rangle-\left\langle
x\left(t_1+\Delta\right)x\left(t_2\right)\right\rangle\right)^2dt_2dt_1.
\end{eqnarray}
The integrand is non-zero only if $t_1+\Delta> t_2$. Introducing the new variable
$\tau=t_2-t_1$ and changing the order of integration, we arrive at the following
expression
\begin{eqnarray}
\nonumber
\mathcal{N}&=&\frac{4}{\left(t-\Delta\right)^2}\int_0^{\Delta}\\
&&\times\int_0^{t-\Delta-\tau}\left(\left\langle x^2\left(t_1+\Delta\right)
\right\rangle-\left\langle x^2\left(t_1+\tau\right)\right\rangle\right)^2dt_1d\tau.
\end{eqnarray}
Introducing the MSD (\ref{x2SBM}) and changing the variable $t_1+\tau_0\to t_1$,
we obtain
\begin{equation}
\mathcal{N}=\frac{16D_0^2\tau_0^2}{\left(t-\Delta\right)^2}\int_0^{\Delta}\int_{
\tau_0}^{\tau_0+t-\Delta-\tau}\log^2\left(\frac{t_1+\Delta}{t_1+\tau}\right)dt_1
d\tau.
\end{equation}
Let us consider the case $\tau_0\ll\Delta\ll t$. Introducing the variables $x=t_1
/\Delta$ and $y=\tau/\Delta$ and changing the upper limit of integration to infinity
and the lower limit to zero in the inner integral, we get
\begin{equation}
\label{XY}
\mathcal{N}=
\frac{16D_0^2\tau_0^2C\Delta^2}{\left(t-\Delta\right)^2}.
\end{equation}
Here the constant $C$ is given by 
\begin{equation}
C=\int_0^1\int_0^{\infty}\log^2\left(\frac{x+1}{x+y}\right)dxdy=\frac{\pi^2}{6}
-1\simeq0.645.
\label{intC}
\end{equation} 
Dividing $\mathcal{N}$ by $\left<\overline{\delta^2(\Delta)}\right>^2$ from
equation (\ref{d1}) we recover the final expression (\ref{eq-eb-log}) for the
ergodicity breaking parameter.

In the case of ageing
\begin{equation}
\mathrm{EB}_a(\Delta)=\lim_{t\to\infty}\frac{\left<\left(\overline{\delta_a^2(
\Delta,t_a)}\right)^2\right>-\left<\overline{\delta_a^2(\Delta,t_a)}\right>^2)}{
\left<\overline{\delta_a^2(\Delta,t_a)}\right>^2}.
\end{equation}
The derivation is similar to the non-aged case,
\begin{eqnarray}
\nonumber
&&\left\langle\left(\overline{\delta_a^2(\Delta,t_a)}\right)^2\right\rangle-
\left\langle\overline{\delta_a^2(\Delta,t_a)}\right\rangle^2\\
&&\hspace*{1.6cm}=\frac{16D_0^2\tau_0^2}{
\left(t-\Delta\right)^2}\int_0^{\Delta}\int_{t_a+\tau_0}^{t_a+\tau_0+t-\Delta-\tau}
\log^2\left(\frac{t_1+\Delta}{t_1+\tau}\right)dt_1d\tau.
\end{eqnarray}
Expanding the integrand for $t_1>t_a\gg 1$, we get
\begin{equation}
\log^2\left(\frac{t_1+\Delta}{t_1+\tau}\right)\simeq\left(\frac{\Delta-\tau}{t_1}
\right)^2.
\end{equation}
Evaluating the integral for $t_a\gg\Delta$, $t\gg\Delta$, we obtain
\begin{equation}
\left<\left(\overline{\delta_a^2(\Delta)}\right)^2\right>-\left<\overline{\delta_a
^2(\Delta)}\right>^2=\frac{16D_0^2\tau_0^2\Delta^3}{3t_at\left(t+t_a\right)},
\end{equation}
and the ergodicity breaking parameter $\mathrm{EB}_a(\Delta)$ is then given by
equation (\ref{EBa}).

\end{document}